\documentclass[11pt,nofootinbib,article]{revtex4}
\usepackage{mathrsfs}
\usepackage{graphicx}
\usepackage{amsmath}
\usepackage{amsfonts}
\usepackage{amssymb}
\usepackage{color}
\usepackage{epsfig}
\usepackage{CJK}
\usepackage{eepic}

\newcommand{\omits}[1]{}

\def\bc{\begin{center}}
\def\nno{\nonumber}
\def\ec{\end{center}}
\def\be{\begin{eqnarray}}
\def\ee{\end{eqnarray}}

\definecolor{dyellow}{rgb}{1.,0.8,.0}
\definecolor{myblue}{rgb}{.1,.1,.7}
\definecolor{dcyan}{rgb}{.0,.6,.6}
\definecolor{cyan}{rgb}{0.4,1.0,1.0}
\definecolor{dmagenta}{rgb}{0.6,0.0,0.6}
\definecolor{brown}{rgb}{0.6,0.2,0.}
\definecolor{darkblue}{rgb}{.0,.0,0.5}
\definecolor{darkred}{rgb}{0.75,0.0,0.0}
\definecolor{orange}{rgb}{1.,.6,.0}
\definecolor{dorange}{rgb}{0.8,.4,.0}
\definecolor{green}{rgb}{0.0,1.0,0.0}
\definecolor{darkgreen}{rgb}{0.0,0.6,0.0}
\definecolor{purple}{rgb}{.4,.0,.4}
\definecolor{lightgrey}{rgb}{0.7, 0.7, 0.7}
\definecolor{grey}{rgb}{0.4, 0.4, 0.4}

\def\ga{\gamma}
\def\dl{\delta}

\def\pa{\partial}

\begin{document}


\title{On Holographic Entanglement Entropy with Second Order Excitations}

\author{Song He$^{2,4,5}$} \email{hesong17@gmail.com}
\author{Jia-Rui Sun$^{1}$} \email{sunjiarui@sysu.edu.cn}
\author{Hai-Qing Zhang$^3$} \email{H.Q.Zhang@uu.nl}

\affiliation{${}^1$Institute of Astronomy and Space Science, Sun Yat-Sen University, Guangzhou 510275, China}
\affiliation{${}^2$ Yukawa Institute for Theoretical Physics, Kyoto University, Kitashirakawa Oiwakecho, Sakyo-ku, Kyoto 606-8502, Japan}
\affiliation{${}^3$ Institute for Theoretical Physics, Utrecht University, Leuvenlaan 4, 3584 CE Utrecht, The Netherlands}
\affiliation{${}^4$ State Key Laboratory of Theoretical Physics, Institute of Theoretical Physics, Chinese Academy of Science, Beijing 100190, China}
\affiliation{${}^5${Max Planck Institute for Gravitational Physics (Albert Einstein Institute)
Am M\"{u}hlenberg 1, 14476 Golm, Germany}}


\begin{abstract}
We study the low-energy corrections to the holographic entanglement entropy (HEE) in the boundary CFT  by perturbing the bulk geometry up to second order excitations. Focusing on the case that the boundary subsystem is a strip, we show that the area of the bulk minimal surface can be expanded in terms of the conserved charges, such as mass, angular momentum and electric charge of the AdS black brane. We also calculate the variation of the energy in the subsystem and verify the validity of the first law-like relation of thermodynamics at second order. Moreover, the HEE is naturally bounded at second order perturbations if the cosmic censorship conjecture for the dual black hole still holds.
\end{abstract}


\maketitle
\newpage
\tableofcontents

\section{Introduction}
The quantum entanglement or entanglement entropy is of great importance in characterizing the correlations between non-local physical quantities in quantum many-body systems. For instance, it can be used as an order parameter to probe the quantum phase transitions at critical points \cite{Osborne:2002,Vidal:2002rm,Calabrese:2004eu,Wen:2006,Kitaev:2005dm}. For a system in pure state with density matrix $\rho$, one can divide it into two spatial regions $A$ and $B$, the entanglement entropy $S_A$ of the subsystem $A$ can be calculated by the von Neumann entropy as
$S_A=-tr_A \left(\rho_A\ln\rho_A\right)$, where $\rho_A=tr_B\rho$ is the reduced density matrix of $A$ obtained by tracing out the degrees of freedom which belong to the Hilbert space of the subsystem $B$. The entanglement entropy can also be described by a more general notion--the entanglement R\'{e}nyi entropy, defined as
$S_n=\frac{1}{1-n}\ln tr\left(\rho_A^n\right)$, where the positive number $n$ is  the order of the R\'{e}nyi entropy, which returns to the von Neumann entropy when taking $n\rightarrow 1$. The important and amazing property of the entanglement entropy is that it is proportional to the area of the boundary surface which divides the systems into subsystems $A$ and $B$, namely, the area law. \footnote{There are indeed some exceptions to violate the area law such as entanglement entropy from fermions \cite{Wolf:2006zzb}.} The area law, together with the fact that entanglement entropy describes the lack of information, say, measured by observer in a subsystem $A$, inspired the studies of interpreting the black hole area entropy as the entanglement entropy between quantum states inside and outside of the horizon \cite{Bombelli:1986rw,Callan:1994py,Holzhey:1994we,Jacobson:1994iw}.

In the framework of the AdS/CFT correspondence or the gauge/gravity duality
\cite{Maldacena:1997re,Gubser:1998bc,Witten:1998qj}, the entanglement entropy also gained a novel holographic interpretation, called the holographic entanglement entropy (HEE) proposed by Ryu and Takayanagi \cite{Ryu:2006bv,Ryu:2006ef}. In this proposal, the entanglement entropy of the subsystem $A$ in the boundary $d$-dimensional CFT can be calculated from the area $A_{\gamma_A}$ of a co-dimensional-2 static minimal surface $\gamma_A$ in the bulk gravity side. More explicitly, the HEE of $A$ is
\be\label{hee}S_A=\frac{A_{\gamma_A}}{4G_{d+1}},\ee
 Soon after, Fursaev gave a proof to the above HEE formulae eq.({\ref{hee}}) by applying the off-shell Euclidean path integral approach to manifolds with conical singularities in the context of the AdS/CFT correspondence \cite{Fursaev:2006ih}. The proof has been improved and generalized from recent studies by Lewkowycz and Maldacena \cite{Lewkowycz:2013nqa} by the replica trick. Besides, another derivation of the HEE for spherical entangling surfaces has been provided by Casini, Huerta and Myers in \cite{Casini:2011kv}. So far there are also many evidences to check the HEE proposal within the AdS$_3$/CFT$_2$ correspondence \cite{Headrick:2010zt,Hartman:2013mia,Faulkner:2013yia}. In addition to various applications of HEE, such as using HEE to probe the confinement/deconfinement phase transition in the large $N$ gauge theories \cite{Klebanov:2007ws} and the phase structures in condensed matter systems \cite{Ogawa:2011bz,Huijse:2011ef,Cai:2012sk,Cai:2012es,Bai:2014tla}, the utility of HEE has been extended to study the renormalization group flows in quantum filed theories \cite{Myers:2012ed,Liu:2012eea,Liu:2013una},  please refer to \cite{Takayanagi:2012kg} for a recent review and references therein.

The original HEE conjecture eq.({\ref{hee}}) was made in the large $N$ and large 't Hooft coupling limit, which indicates that the dual gravitational theory is the classical Einstein theory. In the full version of the AdS/CFT correspondence, due to the strong/weak duality property, there are two kinds of corrections to the theory: one is the higher curvature effects from (super)gravity (which is the $\alpha'$ correction to the boundary CFT), the other is the large $N$ corrections to the boundary quantum field theory, associated with the perturbations from the bulk gravitational theory. For higher curvature or higher derivative terms in the bulk, the corrections to the HEE area law eq.(\ref{hee}) have been studied in the presence of the gravitational Chern-Simons term (with gravitational anomalies) \cite{Sun:2008uf,Castro:2014tta},  and also in quantum field theory which is dual to the general Lovelock gravity in the asymptotically AdS spacetime \cite{deBoer:2011wk,Hung:2011xb,Chen:2013qma,Bhattacharyya:2013jma,Dong:2013qoa,Camps:2013zua,Bhattacharyya:2013gra,
Bhattacharyya:2014yga}. 

The universal behavior of entanglement entropy in low-energy excited states is very important in understanding the quantum entanglement nature of the system, which has been studied by many authors, for example \cite{Alcaraz:2011tn}\cite{Masanes:2009tg}.  Usually it is difficult to investigate the properties of entanglement entropy of systems with generic configurations either from the quantum field theory or from the holographic approaches. However, one can consider two different limits of the subsystem to analyze the properties of the entanglement entropy. The first limit is to study the entanglement entropy with small size for the subsystem, which has been shown in \cite{Alcaraz:2011tn} that the  entanglement entropy and R\'{e}nyi entropy of the 2-dimensional CFT in vacuum and the low-energy excited states are scaled with the primary operators in the theory. Later, the HEE proposal has been applied to study the related problem and the scaling relation for the first order quantum excited entanglement entropy. A novel universal relation between the linearized order entanglement entropy $S^{(1)}$ and the energy $E^{(1)}$ of the subsystem with small size in the boundary CFT has been found \cite{Bhattacharya:2012mi}
\be\label{1stlawee}
E^{(1)} =T_e S^{(1)},
\ee
where $T_e$ is called the entanglement temperature which is proportional to the reciprocal of the size of the CFT. Apparently eq.(\ref{1stlawee}) is reminiscent of the familiar first law of thermodynamics.
For the other limit, one can also consider entanglement entropy of local excited states in which the subsystem is of large size. There are various investigations in this direction from the quantum field theory side \cite{Nozaki:2014hna,He:2014mwa,Nozaki:2014uaa}, in which the authors have proposed that entanglement entropy with large size subsystem is proportional to the
logarithm of quantum dimension of excitations of local operators. The related problems from holographic point of view can also be found in the recent papers \cite{Caputa:2014vaa,Caputa:2014eta,Guo:2013aca,Allahbakhshi:2013rda,Liu:2013iza,Banerjee:2014oaa,Pedraza:2014moa,
Caceres:2014pda,Blanco:2013joa,Wong:2013gua,Nozaki:2013vta,Bhattacharya:2013bna,Lashkari:2013koa,Faulkner:2013ica,Swingle:2014uza,
He:2013rsa,Pang:2013lpa,Barrella:2013wja,Kontoudi:2013rla,Caputa:2013lfa,Caputa:2013eka}.

Despite these interesting progresses, the studies on entanglement entropy with low-energy excitations mostly focused on the linearized perturbations. In the present paper, we will extend the study of the low-energy excitations of the entanglement entropy to second order, from the perspective of the gauge/gravity duality. Specifically, we systematically study the first law-like relation between the variations of entanglement entropy and the variations of energy for the strip subsystem in the boundary CFT up to second order perturbations. We study the HEE and energy in three examples, i.e., general $d$-dimensional (for $d\geq 3$) CFT dual to $d+1$-dimensional Reissner-Nordstr{\"o}m-Anti de Sitter (RN-AdS) black brane, the 2-dimensional CFT dual to 3-dimensional spinning BTZ black hole and the 2-dimensional CFT dual to 3-dimensional non-rotating charged black hole in AdS$_3$. We show that, in the first two examples, the first law-like relation becomes an inequality when taking into account of the second order low-energy corrections (and keeping the spatial size of the subsystem fixed), namely,
\be\label{1stlawee2nd}T_e\left(S^{(1)}+S^{(2)}\right)\leq E^{(1)}+E^{(2)}, \quad{\rm and}\quad S^{(2)}\leq 0,
\ee
where $S^{(i)}$ and $E^{(i)}, (i=1,2)$ are the $i$-th order corrections to entanglement entropy and energy of the subsystem in the CFT. In addition, the cosmic censorship conjecture (CCC) for the black hole will naturally impose an upper bound to the boundary HEE at second order perturbations in each of those three examples. However, in the third example, the inequality (\ref{1stlawee2nd}) can be violated due to the fact that $E^{(2)}=0$, while $S^{(2)}$ is not always negative. This violation is caused by the Weyl anomaly of the $U(1)$ gauge theory in the 2-dimensional CFT. Besides, as was pointed out in \cite{Faulkner:2013ana}, the quantum corrections or loop corrections to the HEE contain two parts, one is from the entanglement between bulk regions separated by the minimal surface, the other one comes from the variations of the minimal surface area. The low-energy corrections considered in this paper is exactly of the second kind. Moreover, we also show that the second order HEE actually reflects the structure of the two-point correlation functions of the boundary renormalized stress tensor. For $d=2$, we show that the corrected scaling relation is consistent with the estimation from the dual CFT$_2$. For other recent studies on second order quantum corrections to entanglement entropy, readers can refer to \cite{Rosenhaus:2014zza}.

The rest parts of the paper are organized as follows: in section \ref{sect:minisurface}, we  describe the general formula for the area functional of bulk co-dimensional-2 surface up to second order perturbations. In sections \ref{sect:general} and \ref{sect:stresstensor}, we study the first law-like relation between HEE and energy for a strip region in the CFT which is dual to RN-AdS$_{d+1}$ black brane. In section \ref{sect:aads3}, we continue studying the HEE with second order excitations for two interesting examples in the asymptotically AdS$_3$ spacetime, one is the spinning BTZ black hole, the other is the charged AdS$_3$ black hole. Conclusions and discussions are drawn in Section \ref{sect:conclusion}.

\section{Perturbations of the bulk co-dimensional-2 surface}\label{sect:minisurface}
The area of the bulk spacelike co-dimensional 2 surface in $d+1$-dimensional spacetime is
\be\label{area} A=\int \sqrt{\gamma} d^{d-1}\xi,
\ee
where
\be\label{worldsheet}
\gamma_{ij}=\frac{\partial X^A}{\partial \xi^i}\frac{\partial X^B}{\partial \xi^j}g_{AB}
\ee
is the induced metric and $\xi^i$ is the coordinate on the surface, while $X^A$ and $g_{AB}$ are the coordinate and the metric of the bulk background spacetime, respectively. Then the variations of the area functional are
\be \label{deltaA}
\delta A &=& -\frac 1 2\int \sqrt{\gamma}\left(\gamma_{ij}\delta\gamma^{ij}+\frac 1 2 \gamma_{ik}\gamma_{jl}\delta \gamma^{ij} \delta \gamma^{kl}+\mathcal{O}\left((\delta\gamma)^3\right)\right)d^{d-1}\xi,\\
\label{delta2A}
\delta^2 A &=& \frac 1 2\int \sqrt{\gamma}\left(\left(\frac 1 2 \gamma_{ij}\gamma_{kl} +\gamma_{ik}\gamma_{jl}\right)\delta\gamma^{ij} \delta \gamma^{kl}+\mathcal{O}\left((\delta\gamma)^3\right)\right)d^{d-1}\xi.
\ee
where $\delta \sqrt{\gamma}=\frac{\sqrt{\gamma}}{2}\left(\gamma^{ij}\delta \gamma_{ij}-\frac 1 2 \gamma^{ik}\gamma^{jl}\delta\gamma_{kj}\delta\gamma_{li}\right)$ (up to the second order). Note that $\sqrt{\gamma}$ and $\gamma_{ij}$ in eqs.(\ref{deltaA}-\ref{delta2A}) are  their zeroth order on-shell values $\sqrt{\gamma^{(0)}}$ and $\gamma^{(0)}_{ij}$. Besides, the variations of $\gamma_{ij}$ and $\gamma^{ij}$ are in the small dimensionless parameter $\varepsilon$ expansion
\be\label{Dgamma} \delta \gamma_{ij}&=&\gamma^{(1)}_{ij}+\gamma^{(2)}_{ij}+\mathcal{O}(\varepsilon^3),\\
\label{DIgamma}
\delta \gamma^{ij}&=& -\gamma^{(1)ij}+\gamma^{(1)ik} \gamma^{(1)j}_k-\gamma^{(2)ij}+\mathcal{O}(\varepsilon^3),\ee
where the indices are lowered and raised by the zeroth order metric $\gamma^{(0)}_{ij}$ and $\gamma^{(0)ij}$. Consequently,
\be\label{area1}
A^{(1)}&=& \frac 1 2\int \sqrt{\gamma^{(0)}}\gamma^{(0)}_{ij}\gamma^{(1)ij}d^{d-1}\xi=\frac 1 2\int \sqrt{\gamma^{(0)}}\gamma^{(1)}d^{d-1}\xi,\\
\label{area2}
A^{(2)}&=& \int \sqrt{\gamma^{(0)}}\left(-\frac 1 2 \gamma^{(1)ij}\gamma^{(1)}_{ij} +\frac 1 8\left(\gamma^{(1)}\right)^2 +\frac 1 2 \gamma^{(2)}\right)d^{d-1}\xi.
\ee
where the induced metric is expanded as
\be\label{gammas}\gamma^{(0)}_{ij}&=&\left(\frac{\partial X^A}{\partial \xi^i}\frac{\partial X^B}{\partial \xi^j}\right)^{(0)}g^{(0)}_{AB},\\
\gamma^{(1)}_{ij}&=&\left(\frac{\partial X^A}{\partial \xi^i}\frac{\partial X^B}{\partial \xi^j}\right)^{(0)}g^{(1)}_{AB}+ \left(\frac{\partial X^A}{\partial \xi^i}\frac{\partial X^B}{\partial \xi^j}\right)^{(1)}g^{(0)}_{AB},\\
\gamma^{(2)}_{ij}&=&\left(\frac{\partial X^A}{\partial \xi^i}\frac{\partial X^B}{\partial \xi^j}\right)^{(0)}g^{(2)}_{AB}+ 2\left(\frac{\partial X^A}{\partial \xi^i}\frac{\partial X^B}{\partial \xi^j}\right)^{(1)}g^{(1)}_{AB}+ \left(\frac{\partial X^A}{\partial \xi^i}\frac{\partial X^B}{\partial \xi^j}\right)^{(2)}g^{(0)}_{AB},
\ee
 The following relation holds if only consider the first order perturbations,
\be
\left(\frac{\partial X^A}{\partial \xi^i}\frac{\partial X^B}{\partial \xi^j}\right)^{(1)}g^{(0)}_{AB}=\delta \left(\frac{\partial X^A}{\partial \xi^i}\frac{\partial X^B}{\partial \xi^j}\right) g^{(0)}_{AB}=0
\ee
provided that the zeroth order EoM of the bulk co-dimensional 2 minimal surface is satisfied (on-shell). However, the above equality does not hold when performing the perturbations up to second order,
\be
\left(\frac{\partial X^A}{\partial \xi^i}\frac{\partial X^B}{\partial \xi^j}\right)^{(1)}g^{(0)}_{AB}\neq 0.
\ee
The asymptotically AdS$_{d+1}$ spacetime in the Poincar\'{e} coordinates is
\be\label{aadsz}
ds^2 &=& \frac{L^2}{\tilde{z}^2}\left(d\tilde{z}^2+ \bar{g}_{\mu\nu}(\tilde{z},x)dx^\mu dx^\nu\right),
\ee
in which $\tilde z$ is the radial coordinate of the AdS$_{d+1}$ spacetime. The general form of the bulk co-dimensional 2 static surface can be expressed by $\tilde z$ as a function of $x^i$ such as $\tilde{z}=\tilde{z}(x^i)$, then the area functional eq.(\ref{area}) becomes
\be\label{areaPoincare}
A=L^{d-1}\int \frac{d^{d-1}x}{\tilde{z}^{d-1}}\sqrt{\det\left(\bar{g}_{ij}(\tilde{z},x)
+\frac{\partial \tilde{z}}{\partial x^i}\frac{\partial \tilde{z}}{\partial x^j}\right)}.
\ee
Let us focus on the situation that eq.(\ref{aadsz}) is a slightly perturbed geometry obtained from the pure AdS spacetime and study the small variation of eq.(\ref{areaPoincare}) which deviates from its pure AdS counterpart. Generally speaking, in order to determine the shape of the bulk minimal surface, one needs to expand $\bar{g}_{ij}=\eta_{ij}+\bar{g}^{(1)}_{ij}+\bar{g}^{(2)}_{ij}$ as well as $\tilde{z}=\tilde{z}^{(0)}+\tilde{z}^{(1)}+\tilde{z}^{(2)}$, and then solve the Euler-Lagrange equation for the bulk static minimal surface up to second order to get $\tilde{z}^{(1)}$ and $\tilde{z}^{(2)}$. However, when the explicit integration form of the area functional (for the bulk minimal surface) in eq.(\ref{areaPoincare}) is known, we can expand it around the minimal surface in pure AdS spacetime in terms of $\varepsilon$ to arbitrary orders. Applying this method, we can study the low-energy excitation corrections to the HEE in vacuum states.

\section{Holographic entanglement entropy in AdS$_{d+1}$ black brane}\label{sect:general}
In the Einstein-Maxwell theory
\begin{equation}
I = \frac{1}{16 \pi G_{d+1}} \int d^{d+1}x \sqrt{-g} \left( R + \frac{d(d-1)}{L^2} - \frac{L^2}{g_s^2} F_{\mu\nu} F^{\mu\nu} \right),
\end{equation}
the charged black brane in AdS$_{d+1}$, i.e. the RN-AdS$_{d+1}$ black brane is ($d\geq 3$)
\be \label{adsbb}
ds^2&=&\frac{L^2}{r^2 f(r)}dr^2+\frac{r^2}{L^2}\left(-f(r)dt^2+dx_i^2\right),\nno\\
f(r) &=& 1 - \frac{M}{r^d} + \frac{Q^2}{r^{2d-2}},\nno
\ee
and
\be
A &=& \mu \left( 1 - \frac{r_o^{d-2}}{r^{d-2}} \right) dt,\quad \mu = \sqrt{\frac{d-1}{2(d-2)}} \frac{g_s Q}{L^2 r_o^{d-2}}.
\ee
where $g_s$ is the dimensionless coupling constant of the $U(1)$ gauge field, $r_o$ is the radius of the outer horizon, $L$ is the curvature radius of the AdS spacetime, $M$ and $Q$ are the mass and charge of the black brane, respectively, and

Besides, the cosmic censorship conjecture (CCC) requires $M\geq 2r_o Q$. In the Poincar\'{e} coordinate $z=L^2/r$, the metric becomes
\be \label{adsbb2}
ds^2=\frac{L^2}{z^2}\bigg(\frac{dz^2}{f(z)}-f(z)dt^2+dx_i^2\bigg),
\ee
with $f(z)=1-\frac{M z^d}{L^{2d}}+\frac{Q^2 z^{2d-2}}{L^{4d-4}}\equiv 1-\tilde{M} \frac{z^d}{L^d}+\tilde{Q}^2\frac{z^{2d-2}}{L^{2d-2}}$, where $\tilde{M}\equiv M/L^d$ and $\tilde{Q}\equiv Q/L^{d-1}$ are dimensionless mass and charge of the RN-AdS$_{d+1}$ black brane.

In order to study the HEE in the background eq.(\ref{adsbb2}) perturbatively, $\tilde{M}$ and $\tilde{Q}$ can be taken as the small parameters which act as the sources of the geometric perturbation away from the pure AdS$_{d+1}$ spacetime (the perturbations can also be caused by adding external matter fields or the fluctuations from the metric itself). Taking $\tilde{M}$ and $\tilde{Q}$ as the first order perturbation, both the metric components $g_{tt}$ and $g_{zz}$ undergo a second order variation as
\be \delta g_{tt}&=& \frac{L^{d-2}}{r^{d-2}}\tilde{M}-\frac{L^{2d-4}}{r^{2d-4}}\tilde{Q}^2
=\frac{z^{d-2}}{L^{d-2}}\tilde{M}-\frac{z^{2d-4}}{L^{2d-4}}\tilde{Q}^2\equiv g^{(1)}_{tt}+ g^{(2)}_{tt},\nno\\
\delta g_{rr}&=& \frac{L^{d+2}}{r^{d+2}}\tilde{M}+\frac{L^{2d+2}}{r^{2d+2}}\tilde{M}^2-\frac{L^{2d}}{r^{2d}}
\tilde{Q}^2=\frac{z^{d+2}}{L^{d+2}}\tilde{M}+\frac{z^{2d+2}}{L^{2d+2}}\tilde{M}^2-\frac{z^{2d}}{L^{2d}}
\tilde{Q}^2\equiv g^{(1)}_{rr}+ g^{(2)}_{rr}.
\ee

We will focus on the case that the boundary CFT is divided into two intervals $A+B$ and taking the subsystem $A$ to be a strip in the region $x_1\in [-l/2, l/2]$ and $x_b\in[-L_0/2, L_0/2]$ with ($b=2,3,\cdots x_{d-1}$). The bulk static minimal surface $\gamma_A$ is a co-dimensional-2 surface, which is described by $x_1=x_1(z)$, then the bulk static hypersurface becomes
\be\label{ss1} ds^2=\frac{L^2}{z^2}\left[\left(\frac{1}{f(z)}+\left(\frac{\partial x_1}{\partial z}\right)^2\right)dz^2+dx_a^2\right],\ee
and its area is
\be\label{areass1} A_{\gamma_A}=2 L^{d-1} L_0^{d-2}\int \frac{dz}{z^{d-1}}\sqrt{\frac{1}{f(z)}+\left(\frac{\partial x_1}{\partial z}\right)^2},\ee
The bulk static minimal surface $\gamma_A$ is determined from the Euler-Lagrange equation, i.e. by minimizing the area functional $A_{\ga_A}$
\be \label{ELx1}\frac{\dl A_{\ga_A}}{\dl x_1}-\pa_z \left(\frac{\dl A_{\ga_A}}{\dl (\pa_z x_1)}\right)=0,\ee
which results in
\be\label{px1} p(z)\equiv\frac{\pa z}{\pa x_1}=\sqrt{f(z)\left(\frac{z_*^{2d-2}}{z^{2d-2}}-1\right)},\ee
where $z_*$ is the turning point of the bulk minimal surface at which $p(z_*)=0$. Besides, it also reflects the embedding of the minimal surface into the bulk which will be affected by the bulk geometric perturbations. Consequently, the area of the bulk static minimal surface becomes
\be \label{minix1} A_{\ga_A}&=& 2L^{d-1}\int_\epsilon^{z_*} dx^{d-2}\frac{dz}{z^{d-1}}\sqrt{\left(\frac{1}{f(z)}+\frac{1}{p(z)^2} \right)}\nno\\
&=&2L^{d-1}\int_\epsilon^{z_*}dx^{d-2}\frac{dz}{z^{d-1}}\frac{1}{\sqrt{f(z)\left(1-\frac{z^{2d-2}}{z_*^{2d-2}}\right)}},\ee
where $\epsilon$ is the geometric short distance cutoff which is related to the UV cutoff $a$ of the dual CFT via the UV/IR relation $\frac{L}{\epsilon}\simeq \frac{L_{A+B}}{a}$, with $L_{A+B}$ the total spatial length of the boundary CFT system $A+B$.

From eqs.(\ref{px1}) (\ref{minix1}) we have
\be\label{px1l} \frac{l}{2}=\int_0^{z_*}\frac{dz}{\sqrt{f(z)\left(\frac{z_*^{2d-2}}{z^{2d-2}}-1\right)}},\ee
where the spatial size $l$ of the subsystem $A$ is fixed. Recall from eq.(\ref{delta2A}) that the variation of the area functional contains the contributions from the bulk metric perturbations as well as the changes in the embedding, both of which are controlled by the source fields, i.e. the conserved charges of the AdS black brane. Therefore, the final expression of the area functional is the expansion in terms of these conserved charges. Defining $\frac{z^{2d-2}}{z^{2d-2}_*}=\xi$, \footnote{It is clear that here $\xi$ is different from $\xi^i$ in section II, which is the coordinates of the induced metric.} then up to order $\mathcal{O}(\tilde{M}^3,\tilde{Q}^3)$ we reach
\be\label{intl} l&=&\frac{z_*}{d-1}\int_0^1 d\xi \left(\xi^{\frac{2-d}{2d-2}}+\frac{\tilde{M} z_*^d}{2L^d}\xi^{\frac{1}{d-1}}+\frac{3\tilde{M}^2 z_*^{2d}}{8L^{2d}} \xi^{\frac{d+2}{2d-2}}-\frac{\tilde{Q}^2 z_*^{2d-2}}{2L^{2d-2}} \xi^{\frac{d}{2d-2}}\right)\left(1-\xi\right)^{-\frac{1}{2}}\nno\\
&=&\frac{z_*\sqrt{\pi}}{d-1}\left(\frac{\Gamma\left(\frac{d}{2d-2}\right)}{\Gamma\left(\frac{2d-1}{2d-2}\right)}
+\frac{\tilde{M} z_*^d}{2L^d}\frac{\Gamma\left(\frac{d}{d-1}\right)}{\Gamma\left(\frac{3d-1}{2d-2}\right)}
+\frac{3\tilde{M}^2 z_*^{2d}}{8L^{2d}}\frac{\Gamma\left(\frac{3d}{2d-2}\right)}{\Gamma\left(\frac{4d-1}{2d-2}\right)}
-\frac{d(d-1)\tilde{Q}^2 z_*^{2d-2}}{(2d-1)L^{2d-2}}\frac{\Gamma\left(\frac{d}{2d-2}\right)}{\Gamma\left(\frac{1}{2d-2}\right)}\right),\quad
\ee
which in turn gives
\be\label{embedding1}z_*=z_*^{(0)}+z_*^{(1)}+z_*^{(2)}+\mathcal{O}(\tilde{M}^3,\tilde{Q}^3),\ee
where
\be\label{embedding2}
z_*^{(0)}&=& \frac{l\Gamma\left(\frac{1}{2d-2}\right)}{2\sqrt{\pi}\Gamma\left(\frac{d}{2d-2}\right)},\nno\\
z_*^{(1)}&=&-\frac{\tilde{M}\sqrt{\pi}}{(d+1)L^d l}\frac{\Gamma\left(\frac{d}{d-1}\right)}{\Gamma\left(\frac{d+1}{2d-2}\right)}
\left(\frac{l\Gamma\left(\frac{1}{2d-2}\right)}{2\sqrt{\pi}\Gamma\left(\frac{d}{2d-2}\right)}\right)^{d+2},\nno\\
z_*^{(2)}&=& \frac{\tilde{M}^2}{L^{2d}}\left(\frac{l\Gamma\left(\frac{1}{2d-2}\right)}{2\sqrt{\pi}\Gamma\left(\frac{d}{2d-2}\right)}\right)^{2d+1}
\left(\frac{1}{4(d+1)}\left(\frac{\Gamma\left(\frac{1}{2d-2}\right)\Gamma\left(\frac{d}{d-1}\right)}{\Gamma\left(\frac{d}{2d-2}\right)\Gamma\left(\frac{d+1}{2d-2}\right)}\right)^2
-\frac{3}{8(2d+1)}\frac{\Gamma\left(\frac{1}{2d-2}\right)\Gamma\left(\frac{3d}{2d-2}\right)}{\Gamma\left(\frac{d}{2d-2}\right)\Gamma\left(\frac{2d+1}{2d-2}\right)}\right)\nno\\
&&+\frac{d\tilde{Q}^2 }{2(2d-1)L^{2d-2}}\left(\frac{l\Gamma\left(\frac{1}{2d-2}\right)}{2\sqrt{\pi}\Gamma\left(\frac{d}{2d-2}\right)}\right)^{2d-1}.
\ee
Therefore, the corresponding area is
\be \label{minix1strip}
 A_{\ga_A}&=& 2L^{d-1} L_0^{d-2}\int_\epsilon^{z_*} \frac{dz}{z^{d-1}}\frac{1}{\sqrt{f(z)\left(1-\frac{z^{2d-2}}{ z_*^{2d-2}}\right)}}\nno\\
 &=&\frac{2 L^{d-1}}{(d-2)}\left(\frac{L_0}{\epsilon}\right)^{d-2}-\frac{2\sqrt{\pi}L^{d-1} L_0^{d-2}}{(d-2)}\frac{\Gamma\left(\frac{d}{2d-2}\right)}{\Gamma\left(\frac{1}{2d-2}\right)}z_*^{2-d}
 +\frac{\tilde{M}}{L^d}\frac{\sqrt{\pi}L^{d-1} L_0^{d-2}}{2(d-1)}\frac{\Gamma\left(\frac{1}{d-1}\right)}{\Gamma\left(\frac{d+1}{2d-2}\right)}z_*^2\nno\\
 &&+\frac{3 \tilde{M}^2}{8L^{2d}}\frac{\sqrt{\pi}L^{d-1} L_0^{d-2}}{(d-1)}\frac{\Gamma\left(\frac{d+2}{2d-2}\right)}{\Gamma\left(\frac{2d+1}{2d-2}\right)}z_*^{d+2}
 -\frac{\tilde{Q}^2L_0^{d-2}\sqrt{\pi}}{L^{d-1}}
 \frac{\Gamma\left(\frac{d}{2d-2}\right)}{\Gamma\left(\frac{1}{2d-2}\right)}z_*^{d}
 +\mathcal{O}(\tilde{M}^3,\tilde{Q}^3).\ee
Substituting eq.(\ref{embedding2}) into eq.(\ref{minix1strip}), we obtain \footnote{Note that eq.(\ref{Astripseries}) holds for $d>2$. The special case is $d=2$, in which the dominant term in the zeroth order in the area is given by $\lim_{\epsilon\rightarrow 0}2L\int_\epsilon^{z_*} \frac{dz}{z}=2L\ln\frac{z_*}{\epsilon}\simeq 2L\ln\frac{l}{a}$, i.e. the logarithmic divergence. }
\be\label{Astripseries}
 A_{\ga_A}^{(0)}&=& \frac{2L^{d-1}}{(d-2)}\left(\frac{L_0}{\epsilon}\right)^{d-2}-\frac{L^{d-1} L_0^{d-2}}{(d-2)l^{d-2}}\left(\frac{2\sqrt{\pi}\Gamma\left(\frac{d}{2d-2}\right)}{\Gamma\left(\frac{1}{2d-2}\right)}\right)^{d-1},\nno\\
A_{\ga_A}^{(1)}&=& \frac{\tilde{M} L_0^{d-2} l^2}{8\sqrt{\pi}(d+1)L}\frac{\Gamma\left(\frac{1}{d-1}\right)}{\Gamma\left(\frac{d+1}{2d-2}\right)}
\left(\frac{\Gamma\left(\frac{1}{2d-2}\right)}{\Gamma\left(\frac{d}{2d-2}\right)}\right)^2,\nno\\
A_{\ga_A}^{(2)}&=& \frac{\tilde{M}^2 L_0^{d-2}\sqrt{\pi} }{4 L^{d+1}} \left(\frac{l\Gamma\left(\frac{1}{2d-2}\right)}{2\sqrt{\pi}\Gamma\left(\frac{d}{2d-2}\right)}\right)^{d+2}
\bigg(\frac{-1}{(d+1)^2(d-1)}\frac{\Gamma\left(\frac{1}{2d-2}\right)}{\Gamma\left(\frac{d}{2d-2}\right)}
\left(\frac{\Gamma\left(\frac{1}{d-1}\right)}{\Gamma\left(\frac{d+1}{2d-2}\right)}\right)^2\nno\\
&&+\frac{3}{2(2d+1)}\frac{\Gamma\left(\frac{d+2}{2d-2}\right)}{\Gamma\left(\frac{2d+1}{2d-2}\right)} \bigg)-\frac{\tilde{Q}^2L_0^{d-2}l}{2L^{d-1}}\left(\frac{l\Gamma\left(\frac{1}{2d-2}\right)}{2\sqrt{\pi}\Gamma\left(\frac{d}{2d-2}\right)}\right)^{d-1}
\ee
Therefore, the corresponding HEE is
\be\label{heestrip}S_{\ga_A}=\frac{1}{4 G_{d+1}}\left(A_{\ga_A}^{(0)}+A_{\ga_A}^{(1)}+A_{\ga_A}^{(2)}\right)+\mathcal{O}(\tilde{M}^3,\tilde{Q}^3).\ee
Note that the first order value is always positive $S_{\ga_A}^{(1)}>0$, while the second order is always negative and it is bounded by the CCC as
\be\label{1st2ndhee}
S_{\ga_A}^{(2)}&\leq & \frac{\tilde{Q}^2 L_0^{d-2}l}{8G_{d+1} L^{d-1}} \left(\frac{l\Gamma\left(\frac{1}{2d-2}\right)}{2\sqrt{\pi}\Gamma\left(\frac{d}{2d-2}\right)}\right)^{d-1}
\bigg(\frac{2r_o^2\sqrt{\pi}}{lL^4} \left(\frac{l\Gamma\left(\frac{1}{2d-2}\right)}{2\sqrt{\pi}\Gamma\left(\frac{d}{2d-2}\right)}\right)^3 \bigg(\frac{3}{2(2d+1)}\frac{\Gamma\left(\frac{d+2}{2d-2}\right)}{\Gamma\left(\frac{2d+1}{2d-2}\right)}\nno\\
&&-\frac{1}{(d+1)^2(d-1)}\frac{\Gamma\left(\frac{1}{2d-2}\right)}{\Gamma\left(\frac{d}{2d-2}\right)}
\left(\frac{\Gamma\left(\frac{1}{d-1}\right)}{\Gamma\left(\frac{d+1}{2d-2}\right)}\right)^2\bigg)-1\bigg).
\ee

\section{The boundary stress tensor of the dual CFT$_d$}\label{sect:stresstensor}
The $d+1$ dimensional bulk spacetime can be written into the ADM form as
\be ds^2=N^2 dr^2+g_{\mu\nu}\left(N^\mu dr+dx^\mu\right)\left(N^\nu dr +dx^\nu\right)\equiv g_{AB}dx^A dx^B
\ee
and the boundary stress tensor can be calculated from the Brown-York formalism
\be\label{stressT} T^{\mu\nu}=\frac{1}{8\pi G_{d+1}}\left(K g^{\mu\nu}-K^{\mu\nu}\right),\ee
in which
\be K_{\mu\nu}=N\Gamma^r_{\mu\nu}=\frac{N}{2}g^{rA}\left(g_{\mu A,\nu}+g_{A \nu,\mu}-g_{\mu\nu,A}\right),
\ee
The variation of the stress tensor with respect to the dimensionless parameter $\varepsilon$ is
\be\label{VstressT} \delta T_{\mu\nu}&=&\frac{-1}{8\pi G_{d+1}}\left(\delta K g_{\mu\nu}+K \delta g_{\mu\nu}-\delta K_{\mu\nu}\right)\nno\\
&=&T^{(1)}_{\mu\nu}+T^{(2)}_{\mu\nu}+\mathcal{O}(\varepsilon^3),
\ee
where
\be
T^{(1)}_{\mu\nu}&=&\frac{-1}{8\pi G_{d+1}}\bigg(K^{(0)}g^{(1)}_{\mu\nu}+K^{(1)}g^{(0)}_{\mu\nu}-K^{(1)}_{\mu\nu}\bigg),\\
T^{(2)}_{\mu\nu}&=&\frac{-1}{8\pi G_{d+1}}\bigg(K^{(0)}g^{(2)}_{\mu\nu}+2K^{(1)}g^{(1)}_{\mu\nu}+\left(K^{(2)}
-K^{(1)}_{\alpha\beta}g^{(1)\alpha\beta}\right)g^{(0)}_{\mu\nu}
-K^{(2)}_{\mu\nu}\bigg),
\ee
with
\be
K^{(1)}&=& g^{(0)\mu\nu}K^{(1)}_{\mu\nu}-g^{(1)\mu\nu}K^{(0)}_{\mu\nu},\nno\\
K^{(2)}&=& g^{(0)\mu\nu}K^{(2)}_{\mu\nu}-g^{(1)\mu\nu}K^{(1)}_{\mu\nu}+g^{(1)\mu\lambda} g^{(1)\nu}_\lambda K^{(0)}_{\mu\nu}-g^{(2)\mu\nu}K^{(0)}_{\mu\nu},\nno\\
K^{(1)}_{\mu\nu}&=& N^{(1)}\Gamma^{(0)r}_{\mu\nu}+N^{(0)}\Gamma^{(1)r}_{\mu\nu},\nno\\
K^{(2)}_{\mu\nu}&=& N^{(2)}\Gamma^{(0)r}_{\mu\nu}+ 2N^{(1)}\Gamma^{(1)r}_{\mu\nu}+N^{(0)}\Gamma^{(2)r}_{\mu\nu}.
\ee
When the boundary is spatially flat, such as the RN-AdS$_{d+1}$ black brane, the boundary counterterm added to cancel the UV divergence is
\be\label{ctterm} I_{\rm ct}=\frac{-(d-1)}{8\pi G_{d+1} L}\int_{\partial \mathcal{M}} d^d x\sqrt{-g},
\ee
which contributes to the boundary stress tensor as
\be\label{Tct} T^{\rm ct}_{\mu\nu}=\frac{2}{\sqrt{-g}}\frac{\delta I_{\rm ct}}{\delta g^{\mu\nu}}=\frac{-(d-1)}{8\pi G_{d+1} L}g_{\mu\nu}.
\ee
Then the first order renormalized stress tensor on the fixed $r$ hypersurface is
\be\label{1stTd}
 T^{(1)}_{\mu\nu}+T^{{\rm ct}(1)}_{\mu\nu} &=& \frac{-1}{8\pi G_{d+1}}\left(-\frac{ g^{(1)}_{\mu\nu}}{L}+\left(\frac{(d-1)r^2}{2L^3}g^{(1)}_{rr}+\frac{g^{(1)\lambda}_\lambda }{L}-\frac{r}{2L}g^{(0)\alpha\beta}g^{(1)}_{\alpha\beta,r}\right)g^{(0)}_{\mu\nu}
+\frac{r}{2L}g^{(1)}_{\mu\nu,r} \right).
\ee
The expectation value of the renormalized stress tensor of the boundary CFT$_d$ can be calculated from variation of the full action with respect to the metric $\bar{g}_{\mu\nu}$ on the conformal boundary (the asymptotic boundary is located at $z=\epsilon\rightarrow 0$) \cite{Balasubramanian:1999re,de Haro:2000xn}.
\be\label{STcft}\delta \left(I+I_{\rm bdy}+I_{\rm ct}\right)= {\rm bulk \quad terms}+\frac{1}{2}\int \sqrt{-\bar{g}}d^d x\langle T_{\mu\nu}\rangle \delta \bar{g}^{\mu\nu},
\ee
where $\bar{g}_{\mu\nu}=\frac{L^2}{r^2}g_{\mu\nu}=\frac{z^2}{L^2}g_{\mu\nu}$, $I_{\rm bdy}$ is the boundary action required by a well-defined variational principle and $\langle T_{\mu\nu}\rangle=\left(\frac{r}{L}\right)^{d-2}\left(T_{\mu\nu}+T^{{\rm ct}}_{\mu\nu}\right)=\left(\frac{L}{z}\right)^{d-2}\left(T_{\mu\nu}+T^{{\rm ct}}_{\mu\nu}\right)$. For the RN-AdS$_{d+1}$ black brane eq.(\ref{adsbb}), the nonvanishing components of $\langle T^{(1)}_{\mu\nu}\rangle$ are
\be\label{1stTRN}\langle T^{(1)}_{tt}\rangle &=&\frac{(d-1)\tilde{M}}{16\pi G_{d+1}L},\nno\\
\langle T^{(1)}_{x_ix_i}\rangle &=&\frac{\tilde{M}}{16\pi G_{d+1}L}.
\ee
Subsequently, the trace of the first order stress tensor is
\be\label{traceT1}\langle T^{(1)\lambda}_{\lambda}\rangle = g^{(0)tt}\langle T^{(1)}_{tt}\rangle+g^{(0)x_ix_i}\langle T^{(1)}_{x_ix_i}\rangle=0,
\ee
which indicates that the boundary dual CFT$_d$ is conformal anomaly free up to first order quantum corrections.

The second order renormalized stress tensor is
\be\label{2ndTd}
 T^{(2)}_{\mu\nu}+T^{{\rm ct}(2)}_{\mu\nu} &=& \frac{-1}{8\pi G_{d+1}}\bigg(-\frac{ g^{(2)}_{\mu\nu}}{L}+\left(\frac{r^2 d}{L^3}g^{(1)}_{rr}+\frac{2g^{(1)\lambda}_\lambda }{L}-\frac{r}{L}g^{(0)\alpha\beta}g^{(1)}_{\alpha\beta,r}\right)g^{(1)}_{\mu\nu}\nno\\
&&+\bigg(\frac{(d-1)r^2}{2L^3}g^{(2)}_{rr}+\frac{(d-1)r^4}{8L^5}\left(g^{(1)}_{rr}\right)^2-\frac{(d-1)g^{(1)r\lambda}g^{(1)}_{r\lambda}}{L}
-\frac{r}{2L}g^{(0)\alpha\beta}g^{(2)}_{\alpha\beta,r}\nno\\
&&-\frac{r^2}{L^3}g^{(1)\lambda}_\lambda g^{(1)}_{rr}+\frac{r}{L}g^{(1)\alpha\beta}g^{(1)}_{\alpha\beta,r}-\frac{g^{(1)\alpha\beta}g^{(1)}_{\alpha\beta}}{L}
+\frac{g^{(2)\lambda}_\lambda }{L}\bigg)
g^{(0)}_{\mu\nu}+\frac{r}{2L}g^{(2)}_{\mu\nu,r}\bigg).
\ee
Thus the nonvanishing components of  $\langle T^{(2)}_{\mu\nu}\rangle$  are
\be\label{2ndT2}
\langle T^{(2)}_{tt}\rangle &=&\frac{-1}{8\pi G_{d+1}}\left(\frac{(3d-11)z^{d}}{8L^{d+1}}\tilde{M}^2
+\frac{(d-1)z^{d-2}}{2L^{d-1}}\tilde{Q}^2\right),\nno\\
\langle T^{(2)}_{x_ix_i}\rangle &=&\frac{-1}{8\pi G_{d+1}}\left(\frac{(11-3d)z^{d}}{8L^{d+1}}\tilde{M}^2
+\frac{(d-1)z^{d-2}}{2L^{d-1}}\tilde{Q}^2\right).
\ee
It is straightforward to check that
\be
\langle T^{(2)\lambda}_{\lambda}\rangle &=&\bar{g}^{(0)\mu\nu}\langle T^{(2)}_{\mu\nu}\rangle-\bar{g}^{(1)\mu\nu}\langle T^{(1)}_{\mu\nu}\rangle\nno\\
&=&\frac{1}{8\pi G_{d+1}}\left(\frac{(3d^2-15d+4)z^{d}}{8L^{d+1}}\tilde{M}^2-\frac{(d-2)(d-1) z^{d-2}}{2L^{d-1}}\tilde{Q}^2\right).
\ee
When taking the UV cutoff to the CFT, i.e., $z=\epsilon\rightarrow 0$ to the asymptotic boundary, $\langle T^{(2)\lambda}_{\lambda}\rangle\rightarrow 0$. With the explicit expressions of $\langle T_{tt}\rangle$, we can calculate the energy associated to the subsystem $A$ in the boundary CFT, which is
\be\label{Ed3}
E&=&\int dx_b^{d-2}\int_{-\frac{l}{2}}^{\frac{l}{2}}dx \langle T_{tt}\rangle\nno\\
&=&E^{(1)}+E^{(2)}+\mathcal{O}(\tilde{M}^3,\tilde{Q}^3),
\ee
where
\be\label{Ed4} E^{(1)}&=&\frac{(d-1)L_0^{d-2}l \tilde{M}}{16\pi G_{d+1}L},\nno\\
E^{(2)}&=&0.
\ee
Therefore, we can check that the first law-like relation for the boundary CFT$_d$ holds at the first order perturbation, namely,
\be\label{1stlawd}T_e S^{(1)}_{\ga_A}=E^{(1)},\ee
with the entanglement temperature given by
\be\label{eT}T_e=\frac{2(d^2-1)\Gamma\left(\frac{d+1}{2d-2}\right)}{\sqrt{\pi}\Gamma\left(\frac{1}{d-1}\right)}
\left(\frac{\Gamma\left(\frac{d}{2d-2}\right)}{\Gamma\left(\frac{1}{2d-2}\right)}\right)^2 l^{-1}.
\ee
When including the contribution from the second order excitations, we have
\be\label{2ndlawd}T_e \left(S^{(1)}_{\ga_A}+S^{(2)}_{\ga_A}\right)<E^{(1)}+E^{(2)}.\ee
Eq.(\ref{2ndlawd}) gives a consistent entropy bound for the subsystem $A$ resembling the Bekenstein bound \cite{Bekenstein:1980jp}. Similar related arguments can be found in the study of the excitation of relative entropy for spherical entangling surface~\cite{Blanco:2013joa}.
\section{Asymptotically AdS$_3$ spacetime}\label{sect:aads3}
In this section, we will study the holographic entanglement entropy of CFT$_2$ with second order excitations in asymptotically AdS$_3$ spacetime, following the spirit that they are formed by small spatially homogeneous metric perturbations from the pure AdS$_3$ spacetime. The first example is the spinning BTZ black hole and the second one is the non-rotating charged black hole in AdS$_3$ spacetime.

\subsection{Spinning BTZ black hole}
The metric of the spinning BTZ black hole \cite{Banados:1992gq,Banados:1992wn} is
\be\label{btz}
ds^2=-\frac{(r^2-r_+^2)(r^2-r_-^2)}{L^2 r^2}dt^2+\frac{L^2 r^2}{(r^2-r_+^2)(r^2-r_-^2)}dr^2
+r^2(d\phi-\frac{r_+r_-}{L r^2}dt)^2, \ee
where the black hole is of mass $M=(r_+^2+r_-^2)/(8G_3 L^2)$, angular
momentum $J=r_+r_-/(4G_3 L)$ and temperature $T=(r_+^2-r_-^2)/(2\pi
r_+ L^2)=\beta^{-1}$. The CCC requires $ML\geq J$, and if we take $ML\equiv \alpha$ to be the small parameter, the magnitude of $J$ cannot be determined generally. In the following we will require that $ML$ and $J$ are of the same order, the special case of which is the near extreme BTZ black hole, namely $ML\rightarrow J$, hence
\be\label{perturbationBTZ}\delta g_{tt}&=&g^{(1)}_{tt}= 8 G_3 M,\quad \delta g_{tx}=g^{(1)}_{tx}=g^{(1)}_{xt}=-\frac{4G_3 J}{L},\nno\\
\delta g_{rr}&=& g^{(1)}_{rr}+g^{(2)}_{rr}=\frac{8G_3 L^4M}{r^4}+\frac{16G_3^2 L^4}{r^6}\left(4M^2L^2-J^2\right),\quad \delta g_{xx}=0.
\ee
So the components of the first and second order renormalized boundary stress tensor are
\be\label{2ndSTbtz}
\langle T^{(1)}_{tt}\rangle &=&\frac{\alpha}{2\pi L^2},\quad \langle T^{(2)}_{tt}\rangle = \frac{G_3}{\pi Lr^2}\left(5M^2L^2+3J^2\right),\nno\\
\langle T^{(1)}_{tx}\rangle &=& \langle T^{(1)}_{xt}\rangle=-\frac{J}{2\pi L},\quad \langle T^{(2)}_{tx}\rangle = \langle T^{(2)}_{xt}\rangle=0,\nno\\
\langle T^{(1)}_{xx}\rangle &=&\frac{\alpha}{2\pi L^2},\quad \langle T^{(2)}_{xx}\rangle = -\frac{G_3}{\pi Lr^2}\left(5M^2L^2+3J^2\right).
\ee
Then $\langle T^{(1)\lambda}_{\lambda}\rangle=0$, and $\langle T^{(2)\lambda}_{\lambda}\rangle=-\frac{14G_3 \alpha^2}{\pi L r^2}-\frac{6G_3J^2}{\pi L r^2}$, which tends to zero at the asymptotic AdS boundary.

The covariant HEE method is required to calculate the HEE for the spinning BTZ black hole, since the background is no longer static. The bulk minimal curve can be easily obtained by converting eq.(\ref{btz}) into the pure Poincar\'{e} coordinate
\be\label{Poincareads3}
ds^2=\frac{L^2}{z^2}\left(dz^2+dw_+dw_-\right)
\ee
via the coordinate transformations
\be
w_\pm &=& \sqrt{\frac{r^2-r_+^2}{r^2-r_-^2}}e^{\left(\phi\pm \frac t L\right)\frac{\Delta_\pm}{L}},\nno\\
z&=&\sqrt{\frac{r_+^2-r_-^2}{r^2-r_-^2}}e^{\frac{\phi r_+}{L}+\frac{t r_-}{L^2}},
\ee
where $\Delta_\pm=r_+\pm r_-$. Note that the HEE of the boundary subsystem $A$ with spatial interval $x\in [-l/2, l/2]$ is simply $S_{\ga_A}=\frac{c}{3}\ln\frac{l}{a}$, which can be explicitly expressed as \cite{Hubeny:2007xt}
\be\label{btzhee}
S_{\ga_A}=\frac{c}{6}\ln\left(\frac{\beta_+\beta_-}{\pi^2 \epsilon^2}\sinh\left(\frac{\pi l}{\beta_+}\right)\sinh\left(\frac{\pi l}{\beta_-}\right)\right),
\ee
in which $c=\frac{3L}{2G_3}$ is the central charge and $\beta_\pm=\frac{2\pi L^2}{r_+\pm r_-}$ are the inverse left and right hand temperatures of the boundary CFT$_2$. Thus, the HEE of the subsystem $A$ can be expanded as
\be \label{heebtz}
S_{\ga_A}&=&S^{(0)}_{\ga_A}+S^{(1)}_{\ga_A}+S^{(2)}_{\ga_A}+\mathcal{O}(\alpha^3,J^3,a^2)\nno\\
&=&\frac{c}{3}\ln\frac{l}{a}+\frac{l^2\alpha}{6L^2}-\frac{l^4}{60c L^4}\left(\alpha^2+J^2\right)+\mathcal{O}(\alpha^3,J^3,a^2)\nno\\
&\leq &\frac{c}{3}\ln\frac{l}{a}+\frac{l^2\alpha}{6L^2}-\frac{l^4 J^2}{30c L^4}+\mathcal{O}(\alpha^3,J^3,a^2)
\ee
where we have used $\alpha\geq J$ in the third line and the relation $\epsilon\simeq a$. Eq.(\ref{heebtz}) shows that the HEE is naturally bounded at second order perturbations if the CCC still holds for the dual BTZ black hole.

The energy of the subsystem $A$ in the boundary CFT$_2$ is
\be\label{Ebtz1}
E&=& \int_{-\frac{l}{2}}^{\frac{l}{2}} dx \langle T_{tt}\rangle\nno\\
&=& E^{(0)}+E^{(1)}+E^{(2)}+\mathcal{O}(\alpha^3,J^3,a^2),
\ee
where
\be\label{2ndEbtz1}
E^{(0)}&=&0,\quad E^{(1)}=\frac{\alpha l}{2\pi L^2}\quad {\rm and}\quad E^{(2)}=0.
\ee
From eqs.(\ref{heebtz})(\ref{2ndEbtz1}) we see that the first order result gives the first law-like relation \cite{Bhattacharya:2012mi} as
\be\label{1stlawbtz}T_e S^{(1)}_{\ga_A}=E^{(1)},\ee
with the entanglement temperature $T_e=\frac{3}{\pi l}$. Again, including the second order excitations, we have the inequality
\be\label{2ndlawbtz}T_e \left(S^{(1)}_{\ga_A}+S^{(2)}_{\ga_A}\right)<E^{(1)}+E^{(2)}.\ee
If $T_e$ is not modified, eq.(\ref{2ndlawbtz}) shows that the subsystem has reached a maximal entropy under small fluctuations.
\omits{But similar to the RN-AdS$_{d+1}$ black brane case, the bulk BTZ black hole is an exact stationary solution. Regarding it as the slightly perturbed geometry from the pure AdS$_3$ spacetime, the perturbations should be stable under all orders of the coupling constants, thus at the second order, the first law-like relation should be
\be\label{2ndlawbtzmod}T'_e \left(S^{(1)}_{\ga_A}+S^{(2)}_{\ga_A}\right)=E^{(1)}+E^{(2)},\ee
with the second order modified entanglement temperature
\be
T'_e=\frac{3}{\pi l}\left(1+\frac{l^2}{10cL^2}\left(\alpha+\frac{J^2}{\alpha}\right)
+\frac{l^4}{100c^2L^4}\left(\alpha+\frac{J^2}{\alpha}\right)^2\right).
\ee
}

Furthermore, recall that the conformal dimension of the boundary stress tensor of the CFT$_2$ dual to the bulk massless graviton is $\Delta=d=2$, so
\be\label{2ndSbtz}
S^{(2)}_{\ga_A}=-\frac{l^4}{60c L^4}\left(M^2L^2+J^2\right)=-\frac{l^{2\Delta}}{60c L^4}\left(M^2L^2+J^2\right),
\ee
which is consistent with the estimation from the dual CFT$_2$ side \cite{Alcaraz:2011tn}.

\subsection{Charged black holes in AdS$_3$}
The charged black hole in AdS$_3$ can be obtained from the Einstein-Maxwell theory
\be\label{3dEM}
I = \frac{1}{16 \pi G_3} \int d^3 x \sqrt{-g} \left( R + \frac{2}{L^2}- \frac{L^2}{ g_s^2}F_{\mu\nu} F^{\mu\nu}\right),
\ee
It has the following metric~\cite{Martinez:1999qi}
\be\label{chargedbtz}ds^2=-\frac{r^2}{L^2}f(r)dt^2
+\frac{L^2}{r^2f(r)}dr^2+r^2d\phi^2,\ee
with
\be
f(r)=1-\frac{M}{r^2}-\frac{Q^2}{4r^2}\ln\left(\frac{r^2}{L^2}\right) \quad {\rm and}\quad A=\frac{g_s Q}{2 L^2}\ln\frac{r}{r_h}dt,
\ee
where $g_s$ is the coupling constant, $Q$ is the charge, and $r_h$ is the black hole outer horizon. When $\frac{M}{L^2}-\frac{Q^2}{4L^2}\left(1-\ln\frac{Q^2}{4L^2}\right)\geq 0$, i.e. the CCC is held, the above metric describes a regular non-rotating charged black hole with radius $L$. In the Poincar\'{e} coordinate eq.(\ref{chargedbtz}) is
\be\label{chargedbtzP}ds^2=\frac{L^2}{z^2}\left(-\left(1-\frac{M z^2}{L^4}+\frac{Q^2z^2}{4L^4}\ln\left(\frac{z^2}{L^2}\right)\right)dt^2
+\frac{dz^2}{\left(1-\frac{M z^2}{L^4}+\frac{Q^2z^2}{4L^4}\ln\left(\frac{z^2}{L^2}\right)\right)}+dx^2\right),\ee
As before, let us treat the mass and charge of the black hole as small perturbations in the pure AdS$_3$ spacetime. When both $M$ and $Q$ are small, the absence of a naked singularity requires $\frac{M}{L^2}$ and $\frac{Q}{L}$ are of the same order of the magnitudes. For convenience, we define $\frac{M}{L^2}\equiv\beta$ and $\frac{Q}{L}\equiv \gamma$. Hence,
\be\label{deltagbtzC}\delta g_{tt}&=&g^{(1)}_{tt}+g^{(2)}_{tt}=\beta+\frac{\gamma^2}{4}\ln\frac{r^2}{L^2},\quad \delta g_{xx}=0,\nno\\
\delta g_{rr}&=& g^{(1)}_{rr}+g^{(2)}_{rr}=\frac{L^4\beta}{r^4}+\left(\frac{L^4\gamma^2}{4r^4}\ln\frac{r^2}{L^2}
+\frac{L^6\beta^2}{r^6}\right).
\ee
Beside, the variation of the boundary holographic stress tensor is
\be\label{2ndSTbtzC}
\langle T^{(1)}_{tt}\rangle &=&\frac{\beta}{16\pi G_3L},\quad \langle T^{(2)}_{tt}\rangle = \frac{5L\beta^2}{64\pi G_3 r^2}+\frac{\gamma^2}{64\pi G_3 L}\ln\frac{r^2}{L^2},\nno\\
\langle T^{(1)}_{tx}\rangle &=& \langle T^{(1)}_{xt}\rangle=0,\quad \langle T^{(2)}_{tx}\rangle = \langle T^{(2)}_{xt}\rangle=0,\nno\\
\langle T^{(1)}_{xx}\rangle &=&\frac{\beta}{16\pi G_3L},\quad
\langle T^{(2)}_{xx}\rangle =-\frac{\gamma^2}{32\pi G_3L}-\frac{5L\beta^2}{64\pi G_3 r^2}+\frac{\gamma^2}{64\pi G_3 L}\ln\frac{r^2}{L^2}.
\ee
Note that both $\langle T^{(2)}_{tt}\rangle$ and $\langle T^{(2)}_{xx}\rangle$ are divergent as $r$ approaches to the boundary. In order to cancel the divergence, the boundary counterterm  from the gauge field is required \footnote{To make the variation of the action in eq.(\ref{3dEM}) be well-defined, the Gibbons-Hawking boundary term and another boundary term for the bulk gauge field which satisfies the Neumann boundary condition should be added, which is $I_N=\frac{L^2}{8\pi G_3 g_s^2}\int\sqrt{-g}dx^2 n^r F_{r\mu}A^\mu$, with $n^r=\frac{r\sqrt{f(r)}}{L}$ to be the unit normal vector of the timelike boundary.}, which is
\be\label{Ictgauge}
I_{\rm ct}^{\rm gauge}= c_1\int dx^2\sqrt{-g}F_{r\mu}F^{r\mu}\ln\frac{r}{L},
\ee
where $c_1$ is the constant which will be fixed by canceling the UV divergence. Eq.(\ref{Ictgauge}) will only contributes to the boundary stress tensor of the CFT$_2$ at the second order,
\be\label{Tgauge}
T^{\rm gauge}_{\mu\nu}=\frac{2}{\sqrt{-\bar{g}}}\frac{\delta I_{\rm ct}^{\rm gauge}}{\delta \bar{g}^{\mu\nu}}=c_1\left(\left(\frac{g_s Q}{2L^2}\right)^2\frac{g_{\mu\nu}}{r^2}+2\frac{r^2f(r)}{L^2}
F_{r\mu}F_{r\nu}\right)\ln\frac{r}{L},
\ee
Therefore, the total second order boundary stress tensor is
\be\label{T2ndtot}
\langle T^{(2)}_{tt}\rangle_{\rm total} &=&\langle T^{(2)}_{tt}\rangle+ T^{{\rm gauge}(2)}_{tt} =\frac{5L\beta^2}{64\pi G_3 r^2}+\frac{\gamma^2}{64\pi G_3 L}\ln\frac{r^2}{L^2}+c_1\frac{g_s^2\gamma^2}{4L^4}\ln\frac{r}{L},\nno\\
\langle T^{(2)}_{xx}\rangle_{\rm total} &=&\langle T^{(2)}_{xx}\rangle+ T^{{\rm gauge}(2)}_{xx} =-\frac{\gamma^2}{32\pi G_3L}-\frac{5L\beta^2}{64\pi G_3 r^2}+\frac{\gamma^2}{64\pi G_3 L}\ln\frac{r^2}{L^2}+c_1\frac{g_s^2\gamma^2}{4L^4}\ln\frac{r}{L},
\ee
which give the finite results to the total stress tensor when choosing $c_1=-
\frac{L^3}{8\pi G_3 g_s^2}$. Finally,
\be\label{T2ndtot2}
\langle T^{(2)}_{tt}\rangle_{\rm total} &=&\frac{5L\beta^2}{64\pi G_3 r^2},\nno\\
\langle T^{(2)}_{xx}\rangle_{\rm total} &=&-\frac{\gamma^2}{32\pi G_3L}-\frac{5L\beta^2}{64\pi G_3 r^2}.
\ee
Therefore, $\langle T^{(1)\lambda}_{\lambda}\rangle=0$, while $\langle T^{(2)\lambda}_{\lambda}\rangle=-\frac{14 L\beta^2}{64\pi G_3 r^2}-\frac{\gamma^2}{32\pi G_3 L}\rightarrow -\frac{\gamma^2}{32\pi G_3 L}$ when $r\rightarrow\infty$, which is the Weyl anomaly of the $U(1)$ vector field in the 2-dimensional CFT~\cite{Jensen:2010em}. The anomaly will contribute an anomalous (a positive) term to the second order entanglement entropy.

In the Poincar\'{e} coordinates, the bulk codimensional-2 surface is $x=x(z)$ and the boundary CFT is divided into two subsystems $A$ and $B$, in which $A$ is located in $x_1\in [-l/2, l/2]$, then
\be\label{btzCsub}
\frac{l}{2}&=&\int_0^{z_*}\frac{dz}{\sqrt{\left(1-\frac{M z^2}{L^4}+\frac{Q^2z^2}{4L^4}\ln\left(\frac{z^2}{L^2}\right)\right)\left(\frac{z_*^2}{z^2}-1\right)}}\nno\\
&=& z_*\left(1+\frac{\beta z_*^2}{3L^2}+\frac{\gamma^2 z_*^2}{36L^2}\left(5-6\ln2+\ln\frac{L}{z_*}\right)+\frac{\beta^2 z_*^4}{5L^4}\right),
\ee
in which $z_*$ is obtained as
\be\label{zstar}z_* &=&\frac{l}{2}-\frac{l^3}{24L^2}\beta+\frac{ l^5}{240L^4}\beta^2
-\frac{l^3\left(5+6\ln\frac{L}{l}\right)}{288L^2}\gamma^2+\mathcal{O}(\beta^3,\beta\gamma^2).
\ee
The area of the bulk minimal curve now is
\be\label{miniAbtzC}
A_{\ga_A}&=& 2L\int_a^{z_*}\frac{dz}{z}\frac{1}{\sqrt{\left(1-\frac{M z^2}{L^4}+\frac{Q^2z^2}{4L^4}\ln\left(\frac{z^2}{L^2}\right)\right)\left(1-\frac{z^2}{z_*^2}\right)}}\nno\\
&=&L\left(2\ln\frac{2z_*}{a}+\frac{\beta}{L^2}z_*^2+\frac{\gamma^2}{2L^2}\left(1-\ln2+\ln\frac{L}{z_*}\right)z_*^2
+\frac{\beta^2}{2L^4}z_*^4\right)+\mathcal{O}(\alpha^3,\beta\gamma^2,a^2)\nno\\
&=&L\left(2\ln\frac{l}{a}+\frac{l^2\beta}{12L^2}-\frac{l^4\beta^2}{1440L^4}
+\left(\frac{l^2}{18L^2}+\frac{l^2}{24L^2}\ln\frac{L}{l}\right)\gamma^2\right)+\mathcal{O}(\alpha^3,\beta\gamma^2,a^2).
\ee
Therefore, the holographic entanglement entropy of the subsystem $A$ is
\be\label{heebtzC}S_{\ga_A}&=&\frac{c}{3}\ln\frac{l}{a}+\frac{cl^2}{72L^2}\beta
-\frac{cl^4}{8640L^4}\beta^2+\frac{cl^2}{36L^2}\left(\frac{1}{3}+\frac{\ln\frac{L}{l}}{4}
\right)\gamma^2+\mathcal{O}(\alpha^3,\beta\gamma^2,a^2),
\ee
in which
\be\label{2ndSbtzC}
S^{(1)}_{\ga_A}&=& \frac{cl^2}{72L^2}\beta,\nno\\
S^{(2)}_{\ga_A}&=&-\frac{cl^4}{8640L^4}\beta^2+\frac{cl^2}{36L^2}\left(\frac{1}{3}+\frac{\ln\frac{L}{l}}{4}
\right)\gamma^2,
\ee
which shows that the second order excitation of HEE from the mass is negative. While the contribution from the charge is positive, in contrast to the $d\geq 3$ charged black brane case considered in Section \ref{sect:general}~\footnote{The positive term in $S^{(2)}_{\ga_A}$ in eq.(\ref{2ndSbtzC}) can be estimated in the off-shell Euclidean path integral approach, see details in Appendix~\ref{app.A}.}. Again, the regular condition (CCC) for the charged AdS$_3$ black hole gives a constraint on its charge and mass, i.e. $\frac{M}{L^2}-\frac{Q^2}{4L^2}\left(1-\ln\frac{Q^2}{4L^2}\right)\geq 0$, which results in the entropy bound for the subsystem $A$
\be\label{2ndbtzCbound}
S^{(2)}_{\ga_A}\leq \frac{cl^2}{36L^2}\left(-\frac{l^2\gamma^4}{3840L^2}\left(1-\ln\frac{\gamma^2}{4}\right)^2
+\left(\frac{1}{3}-\frac{\ln\frac{l}{L}}{4}\right)\gamma^2\right).
\ee
It is worth noticing that, unlike the RN-AdS black brane and the spinning BTZ black hole cases, the function on the right hand side of eq.(\ref{2ndbtzCbound}) is not always negative, although bounded by the CCC. However, it tends to zero for small $l/L$ and $\gamma$, and there exists some region that the right hand side of eq.(\ref{2ndbtzCbound}) becomes negative, namely, $S^{(2)}_{\ga_A}\leq 0$.

In addition, the energy of dual boundary CFT$_2$ is
\be\label{EbtzC}
E&=& \int_{\frac{l}{2}}^{\frac{l}{2}} dx \langle T_{tt}\rangle_{\rm total} \nno\\
&=& E^{(0)}+E^{(1)}+E^{(2)}+\mathcal{O}(\alpha^3,\alpha\gamma^2,a^2),
\ee
where
\be\label{2ndEbtzC}
E^{(0)}=0,\quad E^{(1)}=\frac{ l\beta}{16\pi G_3 L}\quad {\rm and}\quad E^{(2)}=0.
\ee
Therefore, it is straightforward to check that the first law-like relation also holds at first order, i.e.
\be\label{1stlawbtzC}T_e S^{(1)}_{\ga_A}=E^{(1)},
\ee
with $T_e=3/(\pi l)$ the same as in the spinning BTZ black hole. However, when the contributions from the second order excitations are taken into account, we have
\be\label{2ndlawbtzC}T_e S^{(2)}_{\ga_A}\leq \frac{cl}{12\pi L^2}\left(-\frac{l^2\gamma^4}{3840L^2}\left(1-\ln\frac{\gamma^2}{4}\right)^2
+\left(\frac{1}{3}-\frac{\ln\frac{l}{L}}{4}\right)\gamma^2\right),\ee
with $T_e S^{(2)}_{\ga_A}\rightarrow 0$ in the small $l/L$ and $\gamma$ limits, which is different from previous examples. Although the CCC gives an upper bound, it cannot always make $S^{(2)}_{\ga_A}<0$ in the present case. This specific phenomenon is caused by the Weyl anomaly of the background $U(1)$ gauge field in the charged black hole in AdS$_3$ spacetime, which results in the violation of the entropy bound relation $S^{(2)}\leq 0$ at the second order low energy perturbations. A possible explanation is that, since the CCC is given by the relation between the conserved charges of the bulk black hole which is associated with classical symmetries of the theory, it is expected that it will not adequate to constrain the second order entanglement entropy to be negative with gravitational anomaly.

\section{Conclusions and discussions}\label{sect:conclusion}
We studied the HEE of the boundary CFT with low-energy excited states up to second order of the gravitational perturbations (or geometric perturbations) when the spatial region of the boundary subsystem is a strip and examined the first law-like relation at the second order. Our strategy is to start from an exact bulk black brane solution in asymptotically AdS spacetime, and then regard the black brane as the perturbed geometry deviating from its ground state--the pure AdS spacetime, through small fluctuations caused by interactions from external fields or operators, such as the mass and gauge fields. From the viewpoint of the dual boundary CFT, it is equivalent to treat a thermodynamically stable finite temperature CFT (or grand canonical ensemble) with thermal and quantum excitations as the perturbed system deviates from its vacuum state CFT. Following this idea, we solved the bulk co-dimensional-2 minimal surface up to second order in terms of the conserved charges of the AdS black brane, specifically, we obtained the second order low energy excitation corrections to the entanglement entropy of vacuum states from the expansion of the HEE. When the spatial size of the subsystem is fixed, the effective temperature gets no correction and then the first law-like relation of the dual CFT becomes an inequality due to the fact that the second order correction to the energy always vanishes while the corresponding correction to the second order HEE is usually negative. In addition, the HEE is shown to be naturally bounded (an upper bound) at the second order perturbations by the requirement that the CCC holds for the bulk black hole and $S^{(2)}\leq 0$ for the first two examples. An exception appeared in the charged non-rotating black hole in AAdS$_3$ spacetime showed that the second order HEE is not always negative due the Weyl anomaly of the background $U(1)$ gauge field. The anomalous contribution to the HEE can also be seem from the Euclidean form of the renormalized effective action of the theory. The phenomenon observed in the charged AdS$_3$ black hole case indicates that the CCC which is associated with classical symmetries of the bulk theory, is not adequate to constrain the second order HEE $S^{(2)}\leq 0$ in the presence of the gravitational anomaly. Nevertheless, the deep connection between the CCC of the bulk black hole and the entropy bound for the dual CFT requires further study.

We need to emphasize that since all of the perturbations (metric and world-sheet) are known from the starting point, so there is actually ``no'' dynamics for those perturbations in our calculations (for ``no'' dynamics we mean that we don't need to solve the EoMs to obtain the solutions in the present cases). In order to study the second order dynamics of the HEE, we need to consider the fluctuations from the bulk black brane, and then solve the EoMs for the bulk minimal surface and the perturbed Einstein equation at second order.  From the field/operator duality and the structure of the area functional eq.(\ref{area2}), one can see that it includes the contributions from the two-point correlation function of some quasi-local stress tensor, e.g., $\langle T^{(1)}T^{(1)} \rangle$ on the minimal surface. Therefore, it would be interesting to take the bulk minimal surface as an additional boundary for the bulk spacetime and calculate the quasi-local stress tensor on it, we will study this issue in another paper. Moreover, the expansion in terms of the conserved charges is actually related to the Fefferman-Graham expansion, using this property, our method can also be utilized to study the dynamics of HEE with excited states in the asymptotically AdS spacetime.

\section*{Acknowledgement}
We would like to thank Chiang-Mei Chen, Chao-Guang Huang, Yi Ling, Yu Tian, Xiao-Ning Wu and Phil Szepietowski for useful discussions and especially thank Wu-Zhong Guo, Ling-Yan Hung, Jian-Xin Lu, Tadashi Takayanagi for valuable comments. S.H. is supported by JSPS postdoctoral fellowship for foreign researchers and by the National Natural Science Foundation of China (No.~11305235). J.R.S. was supported by the National Natural Science Foundation of China (No.~11205058), the Open Project Program of State Key Laboratory of Theoretical Physics, Institute of Theoretical Physics, Chinese Academy of Sciences, China (No.~Y5KF161CJ1) and the Fundamental Research Funds for the Central Universities. H.Q.Z. was supported in part by the fund of Utrecht University budget associated to Gerard 't Hooft and the Young Scientists Fund of the National Natural Science Foundation of China (No.~11205097).

\begin{appendix}
\section{Estimation of the positive term in second order HEE}\label{app.A}
Recall that the total renormalized effective action $I_{\rm total}$ can be expanded as
\be
I_{\rm total}&=&I^{(0)}_{\rm total}+\frac 1 2\int\sqrt{-\bar{g}} d^2 x \langle T_{\mu\nu}\rangle_{\rm total} \delta\bar{g}^{\mu\nu}+\frac 1 8\int\sqrt{-\bar{g}} d^2 x \int\sqrt{-\bar{g}'} d^2 x'\langle T_{\mu\nu}\rangle_{\rm total}\langle T_{\alpha\beta}\rangle_{\rm total} \delta\bar{g}^{\mu\nu}\delta\bar{g}^{\alpha\beta}\nno\\
&&+\int\sqrt{-\bar{g}} d^2 x \langle J_{\mu}\rangle_{\rm total} \delta A^{(0)\mu}+\frac 1 2\int\sqrt{-\bar{g}} d^2 x \int\sqrt{-\bar{g}'} d^2 x'\langle J_{\mu}\rangle_{\rm total}\langle J_{\nu}\rangle_{\rm total}\delta A^{(0)\mu} \delta A^{(0)\nu}+\cdot\cdot\cdot,\nno\\
\ee
where $I^{(0)}_{\rm total}$ is the zeroth order contribution of $I_{\rm total}$, $A^{(0)}_{\mu}=\frac{g_s Q}{2L^2}\delta^t_\mu$ is the source term of the boundary $U(1)$ gauge field and $\delta A^{(0)}_{\mu}$ is proportional to the expectation value of the boundary current $\langle J_{\mu}\rangle_{\rm total}$, in which
\be
\langle J^{\mu}\rangle_{\rm total}=-\frac{1}{\sqrt{-\bar{g}}}\frac{\delta I_{\rm total}}{\delta A_\mu},
\ee
which gives $\langle J_{t}\rangle_{\rm total}=\frac{Q}{16\pi G_3 g_s L}$ and then $\langle J_{t}\rangle_{\rm total}\delta A^{(0)t}\propto \langle J_{t}\rangle_{\rm total}\langle J^{t}\rangle_{\rm total}=-\left(\frac{Q}{16\pi G_3 g_s L}\right)^2$.

To calculate the entanglement entropy of the boundary CFT, we can apply the off-shell Euclidean path integral approach~\cite{Fursaev:2006ih,Lewkowycz:2013nqa,Sun:2008uf}, the replica trick, i.e., the $n$-copy of the boundary CFT induces a Weyl transformation $\bar{g}_{\mu\nu}\rightarrow n^2\bar{g}_{\mu\nu}\equiv(1+\varepsilon)^2\bar{g}_{\mu\nu}$ on the AdS boundary, where $\varepsilon$ is an infinitesimal real parameter. Then the entanglement entropy of the subsystem $A$ is computed by
\be
S_{\ga_A}=\lim_{\varepsilon\rightarrow 0}\left(1-n\frac{\partial}{\partial n}\right)\left(-I^{\rm E}_{\rm total}\right),
\ee
where the Euclidean version of $I^{\rm E}_{\rm total}=-iI_{\rm total}$ in which $t\rightarrow -i\tau$. It turns out that the term $\langle T^{(2)\lambda}_{\lambda}\rangle$ in eq.(\ref{T2ndtot2}) will contribute a term $\frac{Q^2}{32\pi G_3 L^3}$ to $I^{\rm E}_{\rm total}$. Consequently, it contribute a positive term to the HEE at second order low-energy perturbations.

\end{appendix}



\end{document}